\documentclass[floatfix,twocolumn,aps,prd,showpacs,amsmath,amssymb]{revtex4-1}
\usepackage[latin1]{inputenc}
\usepackage[dvips]{graphicx}

\usepackage{dcolumn}
\usepackage{bm}
\usepackage{dsfont}
\usepackage{color}
\usepackage{ulem} 
\usepackage{url}
\usepackage{amsmath,amssymb}

\newcommand{\PR}[1]{\ensuremath{\left[#1\right]}}
\newcommand{\PC}[1]{\ensuremath{\left(#1\right)}}
\newcommand{\chav}[1]{\ensuremath{\left\{#1\right\}}}

\begin{document}

\title{Excitonic gap generation in thin-film topological insulators}

\author{Nat\'{a}lia Menezes$^{1}$, C. Morais Smith$^{1}$ and Giandomenico Palumbo$^{1}$}
\affiliation{$^1$Institute for Theoretical Physics, Center for Extreme Matter and Emergent Phenomena, Utrecht University, Princetonplein 5, 3584CC Utrecht, the Netherlands}

\date{\today}

\begin{abstract}
In this work, we analyze the excitonic gap generation in the strong-coupling regime of thin films of three-dimensional time-reversal-invariant topological insulators.
 We start by writing down the effective gauge theory in 2+1-dimensions from the projection of the 3+1-dimensional quantum electrodynamics. Within this method, we obtain a short-range interaction, which has the form of a Thirring-like term, and a long-range one. The interaction between the two surface states of the material induces an excitonic gap. By using the large-$N$ approximation in the strong-coupling limit, we find that there is a dynamical mass generation for the excitonic states that preserves time-reversal symmetry and is related to the dynamical chiral-symmetry breaking of our model. This symmetry breaking occurs only for values of the fermion-flavor number smaller than $N_{c}\approx 11.8$. Our results show that the inclusion of the full dynamical interaction strongly modifies the critical number of flavors for the occurrence of exciton condensation, and therefore, cannot be neglected. 
 \end{abstract}

\maketitle

\section{Introduction}
Topological materials are, nowadays, a rich and well developed research field in condensed-matter physics. The study of two-dimensional (2D) topological systems started in the early 80's, with the experimental discovery of the integer quantum Hall effect in GaAs \cite{vonKlitzing}. 
Thereafter, the deep relation between this novel phase and the topological invariant induced by a non-trivial Berry phase was theoretically unveiled \cite{KTNN}. An essential feature of these quantum states is that time-reversal symmetry is broken in the bulk. However, the recent discovery of 2D two-dimensional topological insulators (TIs) \cite{Kane-Mele,Hasan,Bernevig3,Molenkamp1} has opened the way to the exploration and classification of a vast number of novel materials, also in higher dimensions. In 3D, similar versions of 2D TIs have been firstly theoretically formulated \cite{Qi} and then experimentally discovered \cite{Molenkamp,Zhang}.
These systems support surface gapless modes, topologically protected by the non-trivial topological number in the gapped bulk.

Although the free-fermion topological phases have been completely classified for all dimensions in terms of their symmetries \cite{Altland, Ryu}, much less is known about the complete classification and characterization of interacting systems, where a variety of quantum phenomena and quasi-particles emerge in the low-energy regime.
This is the case of anyons in fractional quantum Hall states \cite{Laughlin,Jain,Fradkin} and fractional topological insulators \cite{Goerbig,Murthy-Shankar}, which carry fractional electric charge and spin, Cooper pairs (bound states of spin-up and spin-down electrons) in topological superconductors \cite{Read}, and excitons, i.e. particle-hole bound states in bilayer systems \cite{McDonald, Khveshchenko, Joglekar, Budich, Levitov}.
At the microscopic level, Hubbard-like Hamitonians have been employed in the study of exciton condensation in monolayer \cite{Gamayun} and bilayer graphene \cite{Franz1}, bilayer quantum Hall systems \cite{McDonald,MSmith0, McDonald2} and in 3D thin-film TIs in the class AII \cite{Franz2,Xu,Sokolik}. In the latter case, the electron-hole pairs residing on the surface states can condense to form a topological exciton condensate. This kind of condensation can be seen as an electronic superfluid with dissipationless electronic transport and could enable ultra-low-power and energy-efficient devices, as already proposed in Ref.~\cite{Zaanen}. At a theoretical level, mean-field theory studies show the presence of an excitonic gap induced by the short-range part of the Coulomb interaction between the surface states \cite{Franz2}.

\begin{figure}[!htb]
	\centering
		\includegraphics[width=0.35\textwidth]{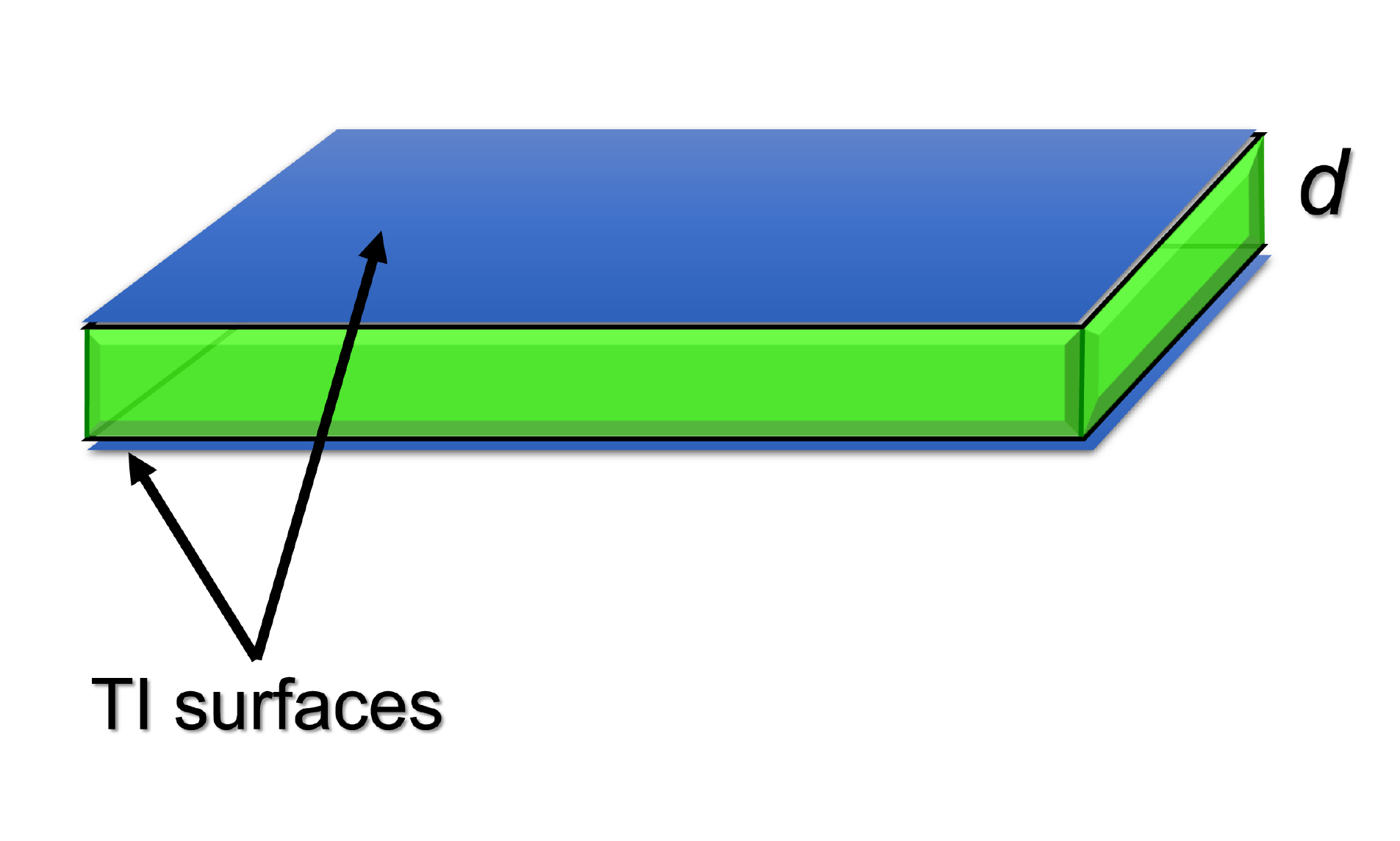}
		\caption{The surfaces of a 3D TI separated by a distance $d$.}
	\label{System}
\end{figure}

In this paper, we propose a precise and self-consistent derivation of the gauge theory describing the short-range interaction in thin films of TIs. In these materials, the free-surface states are defined in terms of massless Dirac fermions and the corresponding interactions are encoded in quantum electrodynamics (QED). Our theoretical model is based on the fact that the massless Dirac fermions are confined on the 2D surfaces, while the virtual photons that mediate their quantum electromagnetic interactions are free to propagate in the 3D surrounding space. This approach has been already successfully employed in the study of several quantum systems, such as graphene \cite{MSmith1,MSmith3}, transition-metal dichalcogenides \cite{MSmith2}, and the edge modes of 2D TIs \cite{Menezes}. The local part of our effective-field theory is given by a generalized 2+1-D Thirring model, which has important applications in both condensed-matter and particle physics \cite{Herbut, Palumbo1, Palumbo2, Gomes}, and represents one the main results of this paper. Importantly, our approach fixes uniquely the value of its coupling constant, which turns out to be proportional to the electric charge and the width of our thin-film TI.

Moreover, if on one hand our work reproduces the effective local Hubbard-like model proposed in Ref.~\cite{Franz2}, on the other hand it does not require any mean-field theory approximation for the identification of the exciton mass gap. By solving the Schwinger-Dyson equation \cite{Schwinger} for the 2+1-D effective field theory in the strong-coupling regime, we show that the mass generation in the exciton condensation is induced dynamically. The dynamical mass generation is due to the breaking of the chiral symmetry \cite{Itoh, Kondo, Gies, Alves}, and represents a non-perturbative phenomenon, beyond the standard mean-field theory.

\section{The model}

We start our analysis with the description of two gapless surface states in 3D thin-film TIs in class AII. They support an odd number of topologically protected helical massless Dirac fermions, which are described by a 2+1-D Dirac theory. We then consider the interactions in and between the two surfaces by including a quantum dynamical U(1) gauge field coupled to the Dirac fermions. This is encoded in the standard QED by introducing a minimal coupling between the gauge potential $A_{\mu}$ and the fermionic current $J_{\mu}$. Importantly, while the masless fermions are confined on the surfaces of the material, the virtual photons that carry the electromagnetic interaction are free to propagate in the 3D space. This is the crucial assumption that will allow us to derive an effective 2+1-D projected theory. Thus, for simplicity, we consider a single Dirac fermion per surface, such that our system is described by the following QED-like action
\begin{eqnarray}
S_{{\rm }}=i\hbar\int d^{3}r\,\PC{\bar{\psi}_{{\rm t}}\bar{\sigma}^{\mu}\partial_{\mu}\psi_{{\rm t}}+\bar{\psi}_{{\rm b}}\sigma^{\mu}\partial_{\mu}\psi_{{\rm b}}}\nonumber \\
-\int d^{4}r\, \left(\frac{\varepsilon_{0}c}{4}\,F_{\alpha\beta}F^{\alpha\beta} + eJ_{3+1}^{\alpha}A_{\alpha}\right), \label{QED4}
\end{eqnarray}
where $\psi_{{\rm b}}$ and $\psi_{{\rm t}}$ denote fermionic fields with $\bar{\psi}_{i}=\psi^{\dagger}_{i}\sigma^{0}$, which are constraint to propagate on the \textit{top} (t) and \textit{bottom} (b) surfaces of the TI, respectively. Here, $\sigma^{\mu}$ are $2\times 2$ Pauli matrices with $\mu=0,1,2$, and we adopt $\bar{\sigma}^{\mu}=-\sigma^{\mu}$, meaning that the two fermions have opposite helicity. The differential elements are given by $d^{3}r=v\,dx\, dy\, dt$ and $d^{4}r=c\, dx\, dy\, dz\, dt$, with $v$ and $c$ the Fermi velocity and the speed of light, respectively. The coupling constant between the matter current and the gauge field $e$ is the electric charge carried by each fermion. $\varepsilon_{0}$ is the vacuum dielectric constant,  $F_{\alpha\beta}=\partial_{\alpha}A_{\beta}-\partial_{\beta}A_{\alpha}$ is the field-strength tensor, $J^{\alpha}_{3+1}=j^{\alpha}_{{\rm t}} +j^{\alpha}_{{\rm b}} =\bar{\psi}_{{\rm t}}\sigma^{\alpha}\psi_{{\rm t}}+\bar{\psi}_{{\rm b}}\sigma^{\alpha}\psi_{{\rm b}}$, and $\alpha,\beta=0,1,2,3$.

We will focus on the interaction between the two fermionic species $\psi_{{\rm t},{\rm b}}$, which in our context represent quasi-particles and quasi-holes confined on two different surfaces. As illustrated in Fig.~\ref{System}, the surfaces of the 3D TI are separated by a distance $d$, which is the width of the thin-film, and we describe the surface Dirac fermions by imposing the following constraints on the matter current
\begin{eqnarray}
j^{\alpha}_{{\rm t},{\rm b}}(t,x,y,z) =
\begin{cases}
j^{\mu}_{{\rm t}}(t,x,y)\delta\PC{z-d/2}, \\
j^{\mu}_{{\rm b}}(t,x,y)\delta\PC{z+d/2}. 
\end{cases}\label{c1}
\end{eqnarray}
Because the fermions interact with a dynamical quantum electromagnetic field, we can integrate out the gauge field to obtain the effective non-local interaction term
\begin{eqnarray}
S^{{\rm eff}}_{{\rm int}} = -\frac{e^{2}}{2}\int d^{4}r d^{4}r'J^{\alpha}_{3+1}(r)\frac{1}{(-\Box)}J_{\alpha}^{3+1}(r'). \label{1}
\end{eqnarray}

By imposing the constraints given in Eq.~(\ref{c1}) we are effectively describing the system as a single surface living in the middle of the thin-film. Hence, Eq.~(\ref{1}) becomes
\begin{eqnarray}
S^{{\rm eff}}_{{\rm int}} &=& -\frac{e^{2}}{2}\int d^{3}r d^{3}r' j^{\mu}_{\kappa}(r)V_{\kappa\rho}(r-r')j_{\mu}^{\rho}(r'), \label{2}
\end{eqnarray}
where $V_{\kappa\rho}(r-r')=\PR{1/(-\Box)}_{\xi_{\kappa\rho}}$, $\kappa,\rho={\rm t},{\rm b}$ and $\xi_{\kappa\rho}$ represents the different values at which the Green's function has to be evaluated. 

Although the system from now on may be treated as an effectively two-dimensional surface, the information about the thin-film width $d$ is carried within the projection. As known in the literature \cite{MacDonald3,Sokolik,Xu}, the exciton condensation in thin-films may only occur when the inter-surface distance $d$ is smaller than an in-plane distance $a$, i.e. $d/a<1$. We introduce this minimal in-plane distance $a$ in our model by shifting the coordinates of the quasiparticles as follows: $r \rightarrow r - a/2$ and $r' \rightarrow r' + a/2$. In this way, Eq.~(\ref{2}) becomes 
\begin{eqnarray}
 S^{{\rm eff}}_{{\rm int}} = -\frac{e^{2}}{2} \int d^{3}r d^{3}r' j^{\mu}_{\kappa}(r-a/2)V_{\kappa\rho}(r-r'-a)j_{\mu}^{\rho}(r'+a/2), \nonumber\\ \label{shift}
\end{eqnarray}
and now the effective interaction carries the information about the length $a$. 

The explicit values of $\xi_{\kappa\rho}$ are
\begin{eqnarray}&&\xi_{{\rm tt}}: z=z'=d/2,\quad \xi_{{\rm tb}}: z=d/2 \ {\rm and}\ z'=-d/2,\nonumber \\
 &&\xi_{{\rm bt}}: z'=d/2 \ {\rm and}\ z=-d/2  ,\quad \xi_{{\rm bb}}: z=z'=-d/2,\nonumber\end{eqnarray}
where, after the projection, the top and bottom components represent two different flavors in the effective middle plane.  
For both $\xi_{{\rm tt}}$ and $\xi_{{\rm bb}}$, we obtain similar results as found in Ref.~\cite{Marino}, namely
\begin{eqnarray}
\PR{\frac{1}{(-\Box)}}_{\xi_{ii}} &=& \frac{1}{2} \int \frac{d^{3}k }{(2\pi)^{3}} \frac{e^{i k\cdot(r-r'-a)}}{\sqrt{k^{2}}}\nonumber\\
&=& \frac{1}{4\pi^{2}(|r-r'-a|^{2}+a^{2})},
\end{eqnarray}
where $a$ settles a minimum distance between the quasiparticles, implying a cutoff on the momenta $k_{{\rm max}} = 1/a$. 
The terms $\xi_{{\rm tb}}$ and $\xi_{{\rm bt}}$ yield
\begin{eqnarray}
\PR{\frac{1}{(-\Box)}}_{\xi_{ij}} =\frac{1}{2} \int \frac{d^{3}k }{(2\pi)^{3}} \frac{e^{-d\sqrt{k^{2}}}e^{i k\cdot(r-r'-a)}}{\sqrt{k^{2}}}. \label{3}
\end{eqnarray}
Now, by considering that $d|k|<1$ \cite{MacDonald3,Sokolik,Gorbar-Miransky}, we expand the exponential $\exp(-d|k|) \approx 1 - d|k|$ and perform the integration over $k$ to find
\begin{eqnarray}
\PR{\frac{1}{(-\Box)}}_{\xi_{ij}} \approx \frac{1}{4\pi^{2}(|r-r'-a|^{2}+a^{2})}- \frac{d}{2}\delta^{3}(r-r'-a).\nonumber\\ \label{4}
\end{eqnarray}
Here, we used the approximation
\begin{eqnarray}
\int \frac{d^{3}k }{(2\pi)^{3}} e^{i k\cdot(r-r'-a)} \approx \delta^{3}(r-r'-a). \label{5}
\end{eqnarray}

We can finally summarize the results for the effective interaction $V_{\kappa \rho}$ after the projection, 
\begin{eqnarray}
&&V_{{\rm tt}}=V_{{\rm bb}}= \frac{1}{4\pi^{2}|r-r'-a|^{2}}, \nonumber \\
&&V_{{\rm tb}}=V_{{\rm bt}}\approx \frac{1}{4\pi^{2}|r-r'-a|^{2}}-\frac{d}{2}\delta(r-r'-a).\nonumber
\end{eqnarray}
where we neglected terms proportional to $a^{2}\approx 0$. By plugging back the interactions above into Eq.~(\ref{shift}), we may write down $S^{{\rm eff}}_{{\rm int}}$ as a long and a short-range contribution (see Appendix A for details).

\section{Single-surface description}

The aim of this section is to describe a two-surface system in terms of a single effective surface with two species of fermions. Our 2+1-D effective action after the projection is given by 
\begin{eqnarray}
S^{{\rm eff}} = i\hbar \int d^{3}r \PC{\bar{\psi}_{{\rm t}}\sigma^{\mu}\partial_{\mu}\psi_{{\rm t}}-\bar{\psi}_{{\rm b}}\sigma^{\mu}\partial_{\mu}\psi_{{\rm b}}} \nonumber\\- \frac{e^{2}}{2\varepsilon_{0}c}\int d^{3}r'\int d^{3}r\  j^{\mu}_{\kappa}\ V_{\kappa \rho}\ j_{\mu}^{\rho}.\label{S3E1}
\end{eqnarray}
where $\kappa,\rho={\rm t},{\rm b}$ represent the different surfaces. Now, we can rewrite the action (\ref{S3E1}) in terms of a single spinor $\Psi = (\psi_{{\rm t}}, \psi_{{\rm b}})^{\top} $. For the kinetic part, we obtain
\begin{eqnarray}
\bar{\psi}_{{\rm t}}\sigma^{\mu}\partial_{\mu} \psi_{{\rm t}} - \bar{\psi}_{{\rm b}}\sigma^{\mu}\partial_{\mu} \psi_{{\rm b}} = \bar{\Psi}\gamma^{\mu}\partial_{\mu}\Psi,\label{S3E2}
\end{eqnarray}
where the $4\times4$ $\gamma$-matrices are defined as \cite{Gomes}
\begin{eqnarray}
\gamma^{\mu}=\left( \begin{array}{cc}
\sigma^{\mu}  &  0 \\
0 &  -\sigma^{\mu}  \end{array} \right) \nonumber,
\end{eqnarray}
with
\begin{eqnarray}
\gamma^{0}=\left( \begin{array}{cc}
\sigma^{0} &  0 \\
0 &  -\sigma^{0} \end{array} \right), \quad  
\gamma^{\tau}=i\left( \begin{array}{cc}
\sigma^{\tau}  &  0 \\
0 &  -\sigma^{\tau}  \end{array} \right). \nonumber
\end{eqnarray}
Here, $\tau=1,2$, $\gamma^{\mu} \equiv \sigma^{0}\otimes\sigma^{\mu}$, and $\otimes$ represents the tensor product. 
The fermionic currents can be written in terms of the new spinors 
\begin{eqnarray}
j^{\mu}_{{\rm t}}&=& \frac{1}{2}\bar{\Psi}(\mathds{1}+\sigma^{0})\otimes\sigma^{\mu}\Psi,  \\ \label{S3E4}
j^{\mu}_{{\rm b}}&=& \frac{1}{2}\bar{\Psi}(\mathds{1}-\sigma^{0})\otimes\bar{\sigma}^{\mu}\Psi, \label{S3E5}
\end{eqnarray}
where $\mathds{1}\otimes\sigma^{\mu}= -i\gamma^{\mu}\gamma^{3}\gamma^{5}$, with
\begin{eqnarray}
\gamma^{3}=i\left( \begin{array}{cc}
0 &  \mathds{1} \\
-\mathds{1} &  0 \end{array} \right), \quad 
\gamma^{5}=i\gamma^{0}\gamma^{1}\gamma^{2}\gamma^{3}=\left( \begin{array}{cc}
0 & \mathds{1} \\
\mathds{1} & 0 \end{array} \right). \nonumber
\end{eqnarray}
Once we have expressed all contributions to the effective action (\ref{S3E1}) in terms of four-component spinors $\bar{\Psi}$ and $\Psi$, we can write down the  following single-surface action
\begin{eqnarray}
&&S^{{\rm eff}}[\bar{\Psi},\Psi] = \frac{e^{2}}{2\varepsilon_{0}c}\int d^{3}r'\int d^{3}r  \mathcal{J}^{\mu}_{35} \frac{1}{4\pi^{2}|r-r'|^{2}} \mathcal{J}_{\mu}^{35}\nonumber\\
&&+ \hbar\int d^{3}r\PR{ i  \bar{\Psi}\gamma^{\mu}\partial_{\mu}\Psi + \frac{e^{2}d}{8\hbar\varepsilon_{0}c} \PC{\mathcal{J}^{\mu}\mathcal{J}_{\mu}+ \mathcal{J}^{\mu}_{35}\mathcal{J}_{\mu}^{35}}},
\label{S3E6}
\end{eqnarray}
where $\mathcal{J}^{\mu} \equiv \bar{\Psi}\gamma^{\mu}\Psi$ and $\mathcal{J}^{\mu}_{35} \equiv \bar{\Psi}\gamma^{\mu}\gamma^{3}\gamma^{5}\Psi$.

\section{Dynamical gap generation}

In the previous section, we derived an effective single-surface interacting model (see Eq.~(\ref{S3E6})), which involves both a short- and a long-range interaction. The former corresponds to a generalized Thirring model \cite{Gies, Palumbo2}, while the latter is similar to the non-local field theory studied in Refs.~\cite{MSmith1, MSmith2,Alves}. These kind of interactions have been already studied separately in the context of dynamical mass generation in Refs.~\cite{Itoh, Kondo,Gies,Alves}. This mechanism is relevant in interacting quantum-field theories and is related to the dynamical breaking of a classical symmetry due to quantum effects. In fact, all the three interaction terms in our effective action (\ref{S3E6}) are invariant under chiral symmetry, which is dynamically broken at the quantum level.
In the first part of this section, we will focus on the short-range interactions $\mathcal{J}^{\mu}\mathcal{J}_{\mu}+\mathcal{J}^{\mu}_{35}\mathcal{J}_{\mu}^{35}$. By following the approach developed in Ref.~\cite{Itoh}, we will show that in the strong-coupling regime both Thirring-like terms yield the same mass generation, and their combined action leads to a larger critical number of fermion flavors $N_{c}$, as compared to a single Thirring term. At last, we will add the long-range interaction and show that the excitonic gap is then enhanced, in agreement with the results found in Refs.~\cite{Gamayun, Alves2} for the case of Gross-Neveu theory.

\subsection{Short-range interactions}

Firstly, let us focus on the dynamical mass generated due to the Thirring-like interactions of Eq.~(\ref{S3E6}). In the large-$N$ approximation, we can write down the effective Lagrangian as
\begin{eqnarray}
&&\mathcal{L}^{{\rm eff}}[\bar{\Psi},\Psi] = i  \hbar\bar{\Psi}_{a}\gamma^{\mu}\partial_{\mu}\Psi_{a}+\nonumber\\
&&  \frac{g}{2N} \PC{\bar{\Psi}_{a}\gamma^{\mu}\gamma^{3}\gamma^{5}\Psi_{a} \bar{\Psi}_{\bar{a}}\gamma_{\mu}\gamma^{3}\gamma^{5}\Psi_{\bar{a}} + \bar{\Psi}_{a}\gamma^{\mu}\Psi_{a} \bar{\Psi}_{\bar{a}}\gamma_{\mu}\Psi_{\bar{a}}},\nonumber
\end{eqnarray}
where $g=e^{2}d N/4\varepsilon_{0}c$. Here the indexes $a,\bar{a}$ denote a sum over $N$ fermion flavors.

Through a Hubbard-Stratonovich transformation, we introduce two auxiliary vector fields $W_{n}^{\mu}$ ($n=1,2$) and two scalar fields $\phi_{n}$ in a way to preserve gauge symmetry. Thus, we obtain
\begin{eqnarray} 
\mathcal{L}^{{\rm eff}} [\bar{\Psi},\Psi,W^{1},W^{2},\phi^{1},\phi^{2}]= i \hbar \bar{\Psi}_{a}\gamma^{\mu}\mathcal{D}_{\mu}\Psi_{a} \nonumber\\
- \sum_{n=1,2}\frac{1}{2g}\PC{W_{n}^{\mu}-\sqrt{N}\partial^{\mu}\phi_{n}}^{2}, \label{S4E4}
\end{eqnarray}
where $\mathcal{D}_{\mu}= \partial_{\mu}-(i/\sqrt{N})\gamma^{3}\gamma^{5}W^{1}_{\mu}-(i/\sqrt{N})W^{2}_{\mu}$.
By following a similar procedure as adopted in Ref.~\cite{Itoh}, we introduce a non-local gauge-fixing term of the form 
\begin{eqnarray}
-\frac{1}{2}\PR{\partial_{\mu}W^{\mu}+\sqrt{N}\frac{\zeta(\partial^{2})}{g}\phi} \frac{1}{\zeta(\partial^{2})} \PR{\partial_{\nu}W^{\nu}+\sqrt{N}\frac{\zeta(\partial^{2})}{g}\phi} \nonumber
\end{eqnarray}
for each gauge field $W^{\mu}_{n}$ in the Lagrangian (\ref{S4E4}). As a result, we obtain
\begin{eqnarray}
&&\mathcal{L}^{{\rm eff}} [\psi, \bar{\psi},W^{1},W^{2}] + \mathcal{L}^{{\rm eff}} [\phi^{1},\phi^{2}] =\nonumber\\
&& i \hbar \bar{\Psi}_{a}\gamma^{\mu}\mathcal{D}_{\mu}\Psi_{a} - \frac{1}{2g}W^{n}_{\mu}W_{n}^{\mu} -\frac{1}{2}\partial_{\mu}W_{n}^{\mu}\frac{1}{\zeta(\partial^{2})}\partial_{\nu}W_{n}^{\nu}\nonumber \\
&&  - \frac{1}{2g}\PR{\zeta(\partial^{2})\phi_{n}}\phi_{n}-\frac{1}{2}\partial_{\mu}\phi_{n}\partial^{\mu}\phi_{n}, \quad\label{S4E6}
\end{eqnarray}
where the gauge-fixing term decoupled the $\phi$-boson fields, which have also been rescaled as $\sqrt{N/g}\phi_{n}  \rightarrow \phi_{n}$. The double index $n$ indicates a summation over the fields. Notice in Eq.~(\ref{S4E6}) that only the strong-coupling regime $g\rightarrow \infty$ preserves gauge symmetry, leading to a massless gauge boson. We shall return to this point later in the Schwinger-Dyson analysis.

Once we have obtained the gauge theory in Eq.~(\ref{S4E6}), we proceed by defining the Feynman rules needed for calculating the mass generation. The full fermion propagator reads
\begin{eqnarray}
S(p)=\frac{i}{A(-p^{2})\gamma^{\mu}p_{\mu}-B(-p^{2})},\label{S4E8}
\end{eqnarray}
where $A$ represents a correction to the fermion-field wave function, and $B$ is the order parameter of the chiral symmetry, which preserves parity in 2+1 dimensions. The Schwinger-Dyson equation for the fermion two-point function is given by
\begin{eqnarray}
S^{-1}(p)= S_{0}^{-1}(p) - i\Sigma (p),\label{S4E9}
\end{eqnarray}
where $S_{0}=i/\gamma^{\mu}p_{\mu}$ is the free-fermion propagator. The self-energy $\Sigma$ contains the contribution from both types of local interaction, and it is determined by
\begin{eqnarray}
-i\Sigma =-\frac{1}{N} \int \frac{d^{3}k}{(2\pi)^{3}} \gamma^{\mu}\gamma^{3}\gamma^{5} S(k)\Gamma^{\nu}\gamma^{3}\gamma^{5} G^{1}_{\mu\nu}(p-k) \nonumber\\
-\frac{1}{N} \int \frac{d^{3}k}{(2\pi)^{3}} \gamma^{\mu} S(k)\Gamma^{\nu} G^{2}_{\mu\nu}(p-k).\qquad \label{S4E10}
\end{eqnarray}
$\Gamma^{\nu}$ and $G^{n}_{\mu\nu}$ are the full-vertex function and the full gauge-boson propagators, respectively. Here, we will adopt the bare-vertex approximation, i.e. $\Gamma^{\nu}=\gamma^{\nu}$. The explicit expression for the full gauge-boson propagator reads
\begin{eqnarray}
G^{n}_{\mu\nu} (k)= iG^{n}_{0}(-k^{2})\PC{g_{\mu\nu}-\eta(-k^{2})\frac{k_{\mu}k_{\nu}}{k^{2}}},\label{S4E14}
\end{eqnarray}
where $G^{1}_{0}=1/(g^{-1}-\Pi)$, $G^{2}_{0}=1/(g^{-1}+\Pi)$, and $\eta$ is a non-trivial function of the momentum related to the non-local gauge approximation \cite{Itoh}. The function $\Pi(-k^{2})$ emerges from the one-loop polarization tensor, inducing dynamics to the gauge fields $W^{n}_{\mu}$ through interaction effects. 

In the strong-coupling regime ($g \rightarrow \infty$), both contributions in Eq.~(\ref{S4E10}) reduce to a single term. By replacing the respective $\Gamma^{\nu}$ and $G^{n}_{\mu\nu}$ functions into Eq.~(\ref{S4E10}) and using that $[\gamma^{\mu},\gamma^{3}\gamma^{5}]=0$, we obtain
\begin{eqnarray}
&&[A(p^{2})-1]\gamma^{\mu}p_{\mu}-B(p^{2}) =\nonumber\\
&& \frac{2}{N}\int \frac{d^{3}k}{(2\pi)^{3}}  \frac{\gamma^{\mu}(A\gamma^{\alpha}k_{\alpha}+B)\gamma^{\nu}}{(A^{2}k^{2}+B^{2})\Pi(q^{2})}  \PC{g_{\mu\nu}-\eta\frac{q_{\mu}q_{\nu}}{q^{2}}},\ 
\label{S4E15}
\end{eqnarray}
where $q=p-k$. We also performed a transformation to the Euclidean space ($k_{0} \rightarrow ik_{0}^{E}$). 

By taking the trace over $\gamma$-matrices in Eq.~(\ref{S4E15}), we obtain two coupled equations: one related to the renormalization of the fermion wavefunction and another related to the generation of the fermionic mass. Within the non-local gauge-fixing picture, the fermion wavefunction is not renormalized. This means that $A(p^2)=1$, and it leads to both
\begin{eqnarray}
0 = \frac{2}{Np^{2}}\int \frac{d^{3}k}{(2\pi)^{3}}  \frac{1}{(k^{2}+B^{2})\Pi} &&\left[(\eta - 1)p\cdot k \right.  \nonumber\\
 &&-\left. 2\eta\frac{(k\cdot q)(p\cdot q)}{q^{2}} \right], \quad
\label{S4E18}
\end{eqnarray}
and
\begin{eqnarray}
B = \frac{2}{N}\int \frac{d^{3}k}{(2\pi)^{3}}  \frac{B \PC{3-\eta}}{(k^{2}+B^{2})\Pi},   \label{S4E19}
\end{eqnarray}
where Eq.~(\ref{S4E18}) is used to determine $\eta(q^{2})$. After some calculations, one finds that in the massless gauge boson limit $g\rightarrow \infty$, $\eta=1/3$ is a constant (see Appendix B for details). Within the Schwinger-Dyson equations, this limit is only defined for a nonzero polarization-tensor contribution, i.e. $\Pi(q^{2}) \neq 0$, as seen in Eq.~(\ref{S4E19}). Hence, the \textit{quenched} approximation $\Pi(q^{2})=0$ sometimes used in the literature \cite{Alves} to simplify calculations can only be used here in the case of a massive gauge boson.

We proceed with the computation by considering the massless gauge boson limit with $\eta=1/3$, which yields
\begin{eqnarray}
B = \frac{128}{3N}\int \frac{d^{3}k}{(2\pi)^{3}}  \frac{B }{(k^{2}+B^{2})\sqrt{(p-k)^{2}}}, \label{S4E26}
\end{eqnarray}
where we used $\Pi(q^{2})= \sqrt{q^{2}}/8$.
The integrals over $k$ in Eq.~(\ref{S4E26}) are performed in spherical coordinates. We first integrate over the solid angle, and then split the remaining integral over positive values of $k$ into two regions,
\begin{eqnarray}
B  &=& \frac{64}{3\pi^{2}N} \left \{ \int_{0}^{p}  dk \frac{k^{2}B(k^{2})}{k^{2}+B^{2}(k^{2})} \frac{1}{|p|} \right. \nonumber \\
&+& \left.  \int_{p}^{\Lambda} dk \frac{k^{2}B(k^{2})}{k^{2}+B^{2}(k^{2})} \frac{1}{|k|} \right\},
\label{S4E27}
\end{eqnarray}
where the virtual-momentum $k$ is, respectively, less or greater than the external momentum $p$. Here, $\Lambda$ is a cutoff and $p=|p|$. Now, we transform the integral Eq.~(\ref{S4E27}) into a differential equation, and by considering $p^{2}+B^{2}(p^{2}) \approx p^{2}$, we obtain 
\begin{eqnarray}
p^{2}\frac{d^{2}B}{dp^{2}}+2p\frac{dB}{dp}+ \frac{64}{3\pi^{2}N} B=0.\label{S4E28}
\end{eqnarray}
The solution of Eq.~(\ref{S4E28}) reads
\begin{eqnarray}
B(p) =\sqrt{\frac{m}{p}}\PR{C_{1}\cos\PC{\lambda \ln \frac{p}{m}} + iC_{2}\sin\PC{\lambda \ln \frac{p}{m}}},\ \quad
\label{S4E29}
\end{eqnarray}
where we have introduced the infrared parameter $m$ such that the ratio $p/m$ is dimensionless and the solution obeys the normalization condition $B(m)=m$. $C_{1}$ and $C_{2}$ are coefficients to be determined according to the ultraviolet (UV) and infrared (IR) boundary conditions. The parameter $\lambda$ indicates the behavior of the solutions of Eq.~(\ref{S4E28}), and it is given by
\begin{eqnarray}
\lambda = \frac{1}{2}\sqrt{\frac{256}{3\pi^{2}N}-1}.\label{S4E30}
\end{eqnarray}
We see in Eq.~(\ref{S4E30}) that there is a critical value $N_{c}=256/3\pi^{2} \approx 8.6$ determining the point at which the solution changes from oscillatory to exponential. This critical number is twice the one in QED$_{2+1}$ with a non-local gauge fixing. For values of $N>256/3\pi^{2}$, the solutions in Eq.~(\ref{S4E29}) are real exponentials, with a contribution that increases in the UV limit. Hence, the only possible solution in this regime is $B(p)=0$ (trivial solution; no mass generation) \cite{Appelquist}. For $N<256/3\pi^{2}$, we obtain the oscillatory solutions (\ref{S4E29}). This implies that $B(p)\neq 0$, and consequently, the chiral symmetry has been broken by the dynamical generation of a fermion mass.

The IR and UV boundary conditions are, respectively,
\begin{eqnarray}
\PR{\frac{dB(p)}{dp}}_{p=m}=0,\ {\rm and} \
\PR{p\frac{dB(p)}{dp}+B(p)}_{p=\Lambda}=0.\ \quad \label{S4E32}
\end{eqnarray}
The IR condition yields a relation between the coefficients $C_{1}$ and $C_{2}$, $C_{1}=2i\lambda C_{2}$. By using this result in the UV condition, we obtain an expression for $m$
\begin{eqnarray}
m = \Lambda \exp\PR{-\frac{1}{\lambda}\arctan\PC{\frac{4\lambda}{4\lambda^{2}-1}}}. \label{S4E33}
\end{eqnarray}
The solution (\ref{S4E29}) can be rewritten as
\begin{eqnarray}
B(p) = m \mathcal{F}\PC{\frac{p}{m},\lambda}, \label{S4E34}
\end{eqnarray}
with
$$\mathcal{F}\PC{\frac{p}{m},\lambda} = \sqrt{\frac{m}{p}} \PR{\cos\PC{\lambda \ln \frac{p}{m}} + \frac{1}{2\lambda}\sin\PC{\lambda \ln \frac{p}{m}}}.$$

So far, we have shown that the Thirring-like interactions derived within the dimensional-reduction method break the chiral symmetry and generate a mass in the fermionic sector with a critical number $N_c$ that is twice the value of the standard Thirring model derived in  Ref.~\cite{Itoh}. This makes sense in the strong-coupling regime because the contributions of both Thirring-like interactions sum up, yielding the multiplicative factor 2 in Eq.~(\ref{S4E15}). 

\subsection{Long-range interaction}

At last, we investigate the effect of the long-range interaction in the strong-coupling regime. First, we rewrite the long-range interaction of Eq.~(\ref{S3E6}) in terms of a gauge theory, e.g.
\begin{eqnarray}
H^{\mu\nu}\frac{1}{\sqrt{\Box}}H_{\mu\nu}+ \bar{g} h_{\mu}\mathcal{J}^{\mu}_{35},\label{S4E35}
\end{eqnarray}
where $H_{\mu\nu}=\partial_{\mu}h_{\nu}-\partial_{\nu}h_{\mu}$ and $\bar{g}$ is the coupling constant. This non-local gauge theory is similar to the one studied in Ref.~\cite{Alves}, where the authors also showed the breaking of chiral symmetry.  

By adding the contribution of the long-range interaction to $\Sigma(p)$ and following a standard procedure, we obtain a differential equation similar to Eq.~(\ref{S4E28}), but with a different coefficient multiplying the fuction $B(p)$. In other words, we obtain a different parameter $\lambda$, namely
\begin{eqnarray}
\lambda' = \frac{1}{2}\sqrt{\frac{4}{N}\PC{\frac{64}{3\pi^{2}}+\frac{8}{\pi^{2}}}-1},\label{S4E36}
\end{eqnarray}
where $32/N\pi^{2}$ is the long-range contribution. The new parameter $\lambda'$ leads to a critical number $N_{c} = 352/3\pi^{2} \approx 11.8$. Thus, the difference between the effect caused by the short- and the long-range interaction is mainly associated to the critical number of fermions (or critical coupling) below which the symmetry is dynamically broken. 

Our results show that the short-range interaction yields the major contribution to the dynamical mass generation when compared to the long-range one. However, both interaction effects add up in a way to increase the value of the critical fermion flavor $N_{c}$ for the occurrence of exciton condensation. This dynamical mechanism is driven mainly by the presence of \textit{electronic interactions} between the surfaces of 3D TI thin-films, and is robust only when the surfaces are strongly interacting. The resulting gap is \textit{time-reversal invariant} and represents a signature of excitonic bound states. 

\subsection{Application: Bi$_{2}$Se$_{3}$ thin-film}

Here, we apply our theoretical results about the dynamical gap generation to Bi$_{2}$Se$_{3}$ thin films. This material is one of the most investigated three-dimensional topological insulators \cite{Zhang,Lu}, together with Bi$_{2}$Te$_{3}$ \cite{Chen}. Experimentally, the size of the gap depends on the material, on the thickness of the film, and on the substrate where the material is grown. In particular, the width of the sample drives the transition from a trivial insulator to a quantum spin Hall insulator, up to the limit in which the material presents the characteristics of a true three-dimensional topological insulator. This transition has been theoretically and experimentally investigated in Ref.~\cite{Zhang}.

In our manuscript, to describe these thin films, we adopted the regime where the distance between the surfaces $d$ -- the width of the 3D TI -- is smaller than the in-plane average separation $a$ between electrons and holes. In general, one would not expect interactions between the surfaces of a 3D TI because of the high values of the bulk dielectric constant. However, the bulk dielectric constant depends on the thickness of the material and decreases for thinner samples \cite{D-Wu,Starkov}. In this limit, the effect of electronic interactions becomes relevant. As we have shown, in the strong coupling regime there is a gap generation in each of the surfaces. 

Within these assumptions, by using Eq.~(\ref{S4E33}) we are able to estimate the excitonic gap generated at zero temperature. This estimative depends on the material and dielectric constant of the substrate via the cutoff $\Lambda$, which in the case of Bi$_2$Se$_3$, for a single Dirac mode ($N=1$), is $0.1$ eV \cite{Sokolik}. By considering these parameters, we theoretically estimate $\lambda\simeq 1.65$ and determine the maximum value for the gap, $m \approx 0.07$ eV, arising from the electronic interactions. Interestingly, this value agrees with the gap measured through ARPES for a thin-film thickness of 4 nm in Bi$_2$Se$_3$ \cite{Zhang}.

\section{Conclusions}

It was theoretically proposed that the excitonic bound states at zero magnetic field may have important technological applications such as for dispersionless switching devices \cite{Application1}, or in the design of topologically protected qubits \cite{Application2}, or in heat exchangers \cite{Zaanen}.
It is also well known that TI-based electronic devices are attractive as platforms for spintronic applications. 
In this work, we provide further theoretical support for exciton condensation in thin-film 3D TIs by investigating the influence of electromagnetic interactions in these systems. 

We started by considering that the photons propagate through the 3D surrounding space where the material is immersed, while the mobile electrons propagate on the two 2D surfaces of the 3D TI. Upon projecting the photon dynamics to these two 2D surfaces, we found the effective intra- and inter-surfaces interaction in the system. The problem was then mapped into a single surface one, in which the top and bottom layers appear as flavors of a single fermionic spinor. Within a single-surface picture, we showed that the fermions interact via two effective short-range and one long-range interaction terms. By using a Hubbard-Stratonovich transformation, we introduced the corresponding effective gauge theory and analyzed the dynamical gap generation through the Schwinger-Dyson equation. This gap term is time-reversal invariant and is associated to the chiral symmetry breaking. 

Our results indicate that the combined effect of short- and long-range interactions that emerge from projecting QED enhance the value of the critical fermion flavor number $N_{c}$ in comparison to models that only include short- or long-range interaction. They also confirm the existence and robustness of excitonic bound states in thin-film TIs in the non-perturbative regime. Notice that these results are achieved in the strongly-coupling regime, which is usually difficult to access with analytic techniques due to the failure of the standard perturbation-theory approach.

The method used here can be extended to multi-layer systems, which involve a larger number of fermion species. This will allow one to analyze the chiral-symmetry breaking and dynamical mass generation in experimentally available samples of multi-layered Dirac materials. At present, the multi-layer samples are of higher quality than the corresponding single-layer ones, and it is therefore essential that theoretical investigation tackle those more complex, multi-flavor systems. Furthermore, the same method can be used to study lower-dimensional excitonic bound states, which have been recently proposed in two parallel nanowires \cite{Nanowire}. This problem will be analyzed in future work.

\acknowledgments
This work was supported by CNPq (Brazil) through the Brazilian government project Science Without Borders. We are grateful to S. Kooi, S. Vandoren, E. C. Marino for fruitful discussions. \\

\appendix 

\begin{widetext}

\section{Effective interactions after projection}

After the projection, we obtain the following interaction terms 
\begin{eqnarray}
&&V_{{\rm tt}}=V_{{\rm bb}}= \frac{1}{4\pi^{2}|r-r'-a|^{2}}, \nonumber \\
&&V_{{\rm tb}}=V_{{\rm bt}}\approx \frac{1}{4\pi^{2}|r-r'-a|^{2}}-\frac{d}{2}\delta(r-r'-a).\nonumber
\end{eqnarray}
where $a^{2}\approx 0$. By plugging back these results into Eq.~(\ref{shift}), we find
\begin{eqnarray}
S^{{\rm eff}}_{{\rm int}} &=& -\frac{e^{2}}{2}\int d^{3}r d^{3}r' j^{\mu}_{t,b}(r-a/2)\frac{1}{4\pi^{2}|r-r'-a|^{2}}j_{\mu}^{t,b}(r'+a/2) \nonumber\\
&-& \frac{e^{2}}{2}\int d^{3}r d^{3}r' j^{\mu}_{t,b}(r-a/2)\PR{\frac{1}{4\pi^{2}|r-r'-a|^{2}}-\frac{d}{2}\delta(r-r'-a)}j_{\mu}^{b,t}(r'+a/2) \nonumber\\
&=& -\underbrace{\frac{e^{2}}{2}\int d^{3}r d^{3}r' j^{\mu}_{t,b}(r-a/2)\frac{1}{4\pi^{2}|r-r'-a|^{2}}j_{\mu}^{t,b}(r'+a/2)}_{r\rightarrow r+a/2;\  r'\rightarrow r'-a/2} \nonumber \\
&-&\underbrace{\frac{e^{2}}{2}\int d^{3}r d^{3}r' j^{\mu}_{t,b}(r-a/2)\frac{1}{4\pi^{2}|r-r'-a|^{2}}j_{\mu}^{b,t}(r'+a/2)}_{r\rightarrow r+a/2;\ r'\rightarrow r'-a/2} + \underbrace{\frac{e^{2}d}{4}\int d^{3}r  j^{\mu}_{t,b}(r+a/2)j_{\mu}^{b,t}(r+a/2)}_{r\rightarrow r-a/2} \nonumber\\
&=& -\frac{e^{2}}{2}\int d^{3}r d^{3}r' j^{\mu}_{t,b}(r)\frac{1}{4\pi^{2}|r-r'|^{2}}j_{\mu}^{t,b}(r') -
\frac{e^{2}}{2}\int d^{3}r d^{3}r' j^{\mu}_{t,b}(r)\frac{1}{4\pi^{2}|r-r'|^{2}}j_{\mu}^{b,t}(r') + \frac{e^{2}d}{4}\int d^{3}r  j^{\mu}_{t,b}(r)j_{\mu}^{b,t}(r). \nonumber\\ \label{6}
\end{eqnarray}

\section{$\eta$-function in the strong coupling regime}

By rewriting Eq.~(\ref{S4E18}) of the main text in spherical coordinates, we obtain
\begin{eqnarray}
&&0 = \frac{1}{Np^{2}}\int_{0}^{\infty} \frac{k^{2}dk}{(2\pi)^{2}}  \frac{1}{k^{2}+B^{2}} \int_{0}^{\pi} d\theta \sin{\theta} \times\nonumber\\
&& \PR{ f_{1}(q^{2},k^{2},p^{2})\cos\theta - f_{2}(q^{2},k^{2},p^{2})\sin^{2}\theta},
\label{B1}
\end{eqnarray}
where
$$f_{1}(q^{2},k^{2},p^{2}) \equiv \tilde{G}_{0}(q^{2})(\eta+1)\sqrt{k^{2}p^{2}},$$
and
$$f_{2}(q^{2},k^{2},p^{2}) \equiv \frac{\tilde{G}_{0}(q^{2}) 2\eta k^{2}p^{2}}{q^{2}}.$$
Here, we denote $\tilde{G}_0 = \lim_{g\rightarrow \infty} G_{0}$, in the massless gauge boson limit. Now, we integrate by parts the first integral over $\theta$ in Eq.~(\ref{B1}), which yields
\begin{eqnarray}
\int_{0}^{\pi} d\theta \sin{\theta}\cos\theta f_{1} = -\int_{0}^{\pi} d\theta \sin^{3}\theta \frac{d\tilde{f}_{1}}{dq^{2}}, \label{B2}
\end{eqnarray}
where we used that $q^{2}=p^{2}+k^{2}-2\sqrt{k^{2}p^{2}}\cos\theta$ and $\tilde{f}_{1}=\sqrt{k^{2}p^{2}}f_{1}$. Replacing the result (\ref{B2}) into Eq.~(\ref{B1}), we find
\begin{eqnarray}
&&0 = \frac{1}{N}\int_{0}^{\infty} \frac{dk}{(2\pi)^{2}} \frac{k^{4}}{k^{2}+B^{2}}  \times\nonumber\\
&&\int_{0}^{\pi} d\theta \sin^{3}{\theta} \chav{ \frac{d[(\eta+1)\tilde{G}_{0}]}{dq^{2}} +\frac{2\eta \tilde{G}_{0}}{q^{2}} }, \label{B3}
\end{eqnarray}
with
\begin{eqnarray}
\frac{d[(\eta+1)\tilde{G}_{0}]}{dq^{2}} +\frac{2\eta \tilde{G}_{0}}{q^{2}} =  \frac{1}{q^{4}}\PR{\frac{d(\eta \tilde{G}_{0}q^{4})}{dq^{2}} + q^{4}\frac{d\tilde{G}_{0}}{dq^{2}} }.\nonumber
\end{eqnarray}

Thus, $\eta$ satisfies the following differential equation
\begin{eqnarray}
d(\eta \tilde{G}_{0}q^{4}) = - q^{4}\frac{d\tilde{G}_{0}}{dq^{2}}dq^{2}, \nonumber
\end{eqnarray}
and
\begin{eqnarray}
\eta(q^{2})&=& \frac{2}{\tilde{G}_{0}(q^{2})q^{4}}\int^{q^{2}}_{0}\tilde{G}_{0}(\zeta^{2}) \zeta^{2}d\zeta^{2} - 1 \nonumber\\ 
&=&\frac{2\Pi(q^{2})}{q^{4}}\int^{q^{2}}_{0} \frac{\zeta^{2}}{\Pi(\zeta^{2})}d\zeta^{2} - 1 =\frac{1}{3} ,\label{B4}
\end{eqnarray}
where $\Pi(q^{2}) = \sqrt{q^{2}}/8$. 

\end{widetext}

\end{document}